\documentclass[5p,times,twocolumn,nopreprintline]{elsarticle}

\usepackage{12be-shape}
\usepackage{hyperref}
\usepackage{braket}
\newcommand{\intr}{{\mathrm{intr}}}
\newcommand{\nrm}{{\mathrm{norm}}}
\newcommand{\thetaexp}{\theta^{\mathrm{exp}}_{0^+}}

\begin{document}
\ifproofpre{}{\count\footins = 1000}

\title{Intruder band mixing in an \emph{ab initio} description of $\isotope[12]{Be}$ }

\author[1,2,3,4]{A. E. McCoy\corref{em}\orcidlink{0000-0002-1033-1474}}
\ead{amccoy@anl.gov}
\author[5]{Mark A. Caprio\,\orcidlink{0000-0001-5138-3740}}
\author[6]{Pieter Maris\,\orcidlink{0000-0002-1351-7098}}
\author[1,5]{Patrick J. Fasano\,\orcidlink{0000-0003-2457-4976}}
\cortext[em]{Corresponding author.}
\affiliation[1]{
	organization={Physics Division, Argonne National Laboratory},
	city={Argonne},
	state={Illinois},
	postcode={60439},
	country={USA}
	}
\affiliation[2]{
	organization={Institute for Nuclear Theory, University of Washington},
	city={Seattle},
	state={Washington},
	postcode={98195},
	country={USA}
	}
\affiliation[3]{
	organization={Facility for Rare Isotope Beams, Michigan State University},
	city={East Lansing},
	state={Michigan},
	postcode={48824},
	country={USA}
	}
\affiliation[4]{
	organization={Washington University in Saint Louis}, 
	city={Saint Louis},
	state={Missouri},
	postcode={63130},
	country={USA}
	}
\affiliation[5]{
	organization={Department of Physics and Astronomy, University of Notre Dame},
	city={Notre Dame},
	state={Indiana},
	postcode={46556-5670},
	country={USA}
	}
\affiliation[6]{
	organization={Department of Physics and Astronomy, Iowa State University},
	city={Ames},
	state={Iowa},
	postcode={50011-3160},
	country={USA}
	}

\date{\today}

\begin{abstract}
The spectrum of $\isotope[12]{Be}$ exhibits exotic features, \textit{e.g.}, an intruder ground state and shape coexistence, normally associated with the breakdown of a shell closure.  While previous phenomenological treatments indicated the ground state has substantial contributions from intruder configurations, it is only with advances in computational abilities and improved interactions that this intruder mixing is observed in \emph{ab initio} no-core shell model (NCSM) predictions.  In this work, we extract electromagnetic observables and symmetry decompositions from the NCSM wave functions to demonstrate that the low-lying positive parity spectrum can be explained in terms of mixing of rotational bands with very different intrinsic structure coexisting within the low-lying spectrum. These observed bands exhibit an approximate $\grpsu{3}$ symmetry and are qualitatively consistent with Elliott model predictions.
\relax
\end{abstract}

\maketitle

\section{Introduction}
A breakdown of the $N=8$ shell closure
in neutron-rich $\isotope[12]{Be}$ is supported by both experimental~\cite{
npa-968-2017-71-Kelley, prc-18-1978-2727-Alburger,plb-206-1988-592-Tanihata,prc-48-1993-R1484-Zahar,prc-50-1994-1355-Fortune,prl-82-1999-1383-Freer, plb-491-2000-8-Iwasaki,plb-481-2000-7-Iwasaki,prl-96-2006-032502-Pain, prl-108-2012-122501-Meharchand, prl-85-2000-266-Navin,
plb-560-2003-31-Shimoura,npa-738-2004-333-Bohlen,plb-654-2007-87-Shimoura,ijmpe-17-2008-2067-Bohlen,plb-673-2009-179-Imai,plb-682-2010-391-Kanungo,prc-83-2011-057304-Peters, prl-108-2012-142501-Krieger, npa-875-2012-8-Ilieva,prc-88-2013-044619-Johansen, plb-780-2018-227-Morse,prc-99-2019-064610-Lyu} and theoretical~\cite{fbs-29-2000-131-Baye,jpcs-445-2013-012035-Maris,prc-56-1997-847-Suzuki, prc-69-2004-041302R-Gori,jpg-2-1976-L45-Barker,prc-60-1999-064323-Sherr,npa-703-2002-593-Nunes,plb-505-2001-71-Descouvemont,jpg-34-2007-2715-Hamamoto,plb-660-2008-32-Romero-Redondo,prc-53-1996-708-Thompson, prc-62-2000-034301-Itagaki,prc-66-2002-R011303-Kanada-Enyo,prc-85-2012-014302-Ito,prc-77-2008-054313-Romero-Redondo,prc-68-2003-014319-Kanada-Enyo,npa-836-2010-242-Dufour,prc-91-2015-014310-Maris,prl-100-2008-182502-Ito,prc-100-2019-064616-Kanada-Enyo,ptps-142-2001-205-Kanada-Enyo}
evidence.  The compressed $0_1^+$ to $2_1^+$ level
spacing~\cite{npa-968-2017-71-Kelley,npa-745-2004-155-Tilley},
larger proton radius~\cite{prl-108-2012-142501-Krieger}, and higher $B(E2;
2_1^+\rightarrow 0_1^+)$
value~\cite{plb-673-2009-179-Imai,plb-780-2018-227-Morse}, in comparison with
neighboring $\isotope[10]{Be}$, all point to a significantly deformed intrinsic
state, not the spherical shape expected at a shell closure.  Shell model and
cluster molecular orbital descriptions, as well as spectroscopic factors, all
indicate the $\isotope[12]{Be}$ ground state is an admixture of $0\hbar\omega$ (normal)
configurations with a filled neutron $p$ shell and $2\hbar\omega$ (intruder) configurations with
two neutrons promoted to the $sd$
shell~\cite{jpg-2-1976-L45-Barker,prc-56-1997-847-Suzuki,prl-96-2006-032502-Pain,plb-481-2000-7-Iwasaki,prl-108-2012-122501-Meharchand,prc-69-2004-041302R-Gori,prl-85-2000-266-Navin}.
Although the specific admixture is model dependent, the largest contribution is
consistently identified to be $2\hbar\omega$, \textit{i.e.}, the ground state is an
intruder state.

Efforts to understand the structure of these low-lying
states in $\isotope[12]{Be}$ have largely been restricted to phenomenological
models.  In this work, we focus on understanding the low-lying spectrum of
$\isotope[12]{Be}$ from an \emph{ab initio} perspective.  Specifically, we use the no-core shell model (NCSM) framework~\cite{ppnp-69-2013-131-Barrett} in which energies and wave functions are obtained by solving the non-relativistic Schr\"odinger equation in a basis of antisymmetrized products of harmonic oscillator states\footnote{The NCSM results presented in this work were obtained using the code MFDn~\cite{ccpe-10-2013-Aktulga,hpc-2016-336-Cook,tpds-28-2017-1550-Aktulga}.
}.
Early \emph{ab initio} calculations of $\isotope[12]{Be}$ were
computationally limited to model spaces insufficient to reproduce the intruder
nature of the ground
state~\cite{npa-836-2010-242-Dufour,prc-91-2015-014310-Maris}.  However, here, by
combining computational advances with
use of the Daejeon16 internucleon interaction~\cite{plb-761-2016-87-Shirokov} based on chiral effective field theory but softened to improve numerical precision in a truncated basis, we obtain a
calculated spectrum which is in reasonable agreement with experiment.

\emph{Ab initio} calculations thus obtained~--- without explicit inclusion of
shell structure, clustering, or collectivity~--- can now be used to probe the
intrinsic structure of the low-lying spectrum of $\isotope[12]{Be}$ and identify simple, more intuitive pictures for approximately describing the spectrum.  In particular, we focus on the intrinsic structure of the ground state and the long lived $0^+$ state at 2.251(1)~MeV.

Intruder states are thought to be a result of competition between
shell structure and particle correlations
\cite{ppnp-120-2021-103866-Nowacki,rmp-83-2011-1467-Heyde,ppnp-61-2008-602-Sorlin,prc-41-1990-1147-Warburton}.  Near a shell closure, normal
configurations have little correlation energy, while intruder configurations can achieve a much larger correlation energy
through deformation.  Thus intruder states are expected to have highly deformed intrinsic states relative to other nearby normal states (shape coexistence), which often results in intruder rotational bands with large moments of inertia.   In the low-lying NCSM-calculated spectrum presented in this work, such intruder bands  are observed along with a normal band built on the excited $0^+$ state.  Similar bands were observed in previous theoretical investigations~\cite{prc-91-2015-014310-Maris,prc-68-2003-014319-Kanada-Enyo}.   In this work, the intrinsic shapes of these bands are probed by calculating proton and neutron radii and quadrupole moments.

However, the simple description of shape-coexistent rotational bands is insufficient to describe the low-lying spectrum of $\isotope[12]{Be}$.  Transitions between bands with markedly different intrinsic shape are expected to vanish~\cite{prc-37-1988-2170-Heyde,npa-651-1999-323-Wood}.  Thus the measured $0^+_2\rightarrow0^+_1$ $E0$ transition~\cite{plb-654-2007-87-Shimoura} can be taken as an indication of mixing~\cite{prc-37-1988-2170-Heyde}.  To gain insight into the mixing of the states as well as extract properties of the pure rotational bands, we apply a two-state mixing analysis to the calculated spectrum.  We demonstrate that two-state mixing combined with the rotational picture well describes the low-lying spectrum of $\isotope[12]{Be}$.

In light nuclei, where intrinsic shape is often not sharply defined, the assumption of vanishing transitions between states with different shape is not as well motivated as in heavier systems.  However, as we demonstrate in this work, the vanishing interband transitions can alternatively be understood in the context of an emergent approximate symmetry, specifically Elliott's $\grpsu{3}$ symmetry~\cite{prsla-245-1958-128-Elliott,prsla-245-1958-562-Elliott,prsla-272-1963-557-Elliott,ppnp-67-2012-516-Draayer,prl-111-2013-252501-Dytrych,cpc-207-2016-202-Dytrych,ppnp-89-2016-101-Launey,McCoy2018,prl-125-2020-102505-McCoy,epja-56-2020-120-Caprio,jpg-48-2021-075102-Zbikowski,bjp-49-2022-057066-Caprio} which is tied to both nuclear rotation and deformation as well as microscopic correlations.   We will demonstrate that the rotational bands exhibit this approximate symmetry and discuss the consequences for transition strengths.

In this work, we first present the NCSM calculated spectrum and identify emergent rotational
bands (Sec.~\ref{sec:ncsm}).  We then apply the two-state mixing model to ``un-mix'' the rotational bands and extract information about the intrinsic states of the pure rotational bands (Sec.~\ref{sec:mixing}).   Finally, we interpret the NCSM results in the context of Elliott's $\grpsu{3}$ framework (Sec.~\ref{sec:elliott-su3}).

 \section{Intruder band in $\isotope[12]{Be}$}
\label{sec:ncsm}
\begin{figure}[t]
  \begin{center}
    \includegraphics[width=\columnwidth]{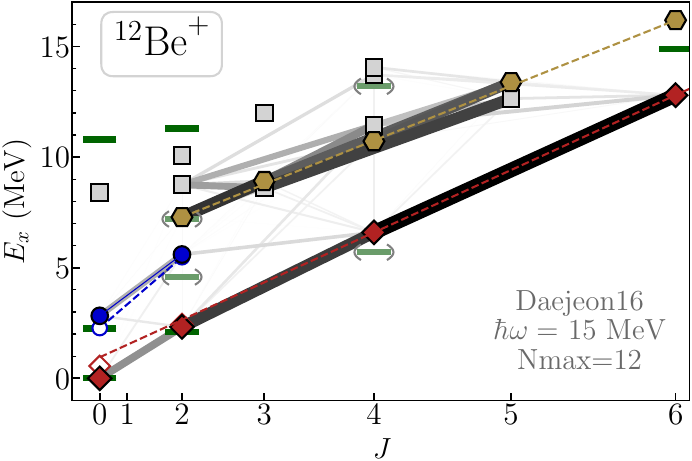}
  \end{center}
  \caption{\label{fig:network} Calculated spectrum of $\isotope[12]{Be}$ with $\hbar\omega=15~\mathrm{MeV}$ at $\Nmax=12$.  The $J$ axis is scaled by $J(J+1)$ but labeled by $J$.  Lines connecting states denote $E2$ transitions.  Thickness and shading correspond to transition strength.  Horizontal green lines indicate experimental energies~\cite{npa-968-2017-71-Kelley}. Parentheses indicate $J^\pi$  assignment is only tentative.  Open symbols indicate excitation energies of pure states (see Sec.~\ref{sec:mixing}). Dashed lines indicate rotational energy fit to pure band members.}
\end{figure}
\begin{figure}[tb]
  \begin{center}
    \includegraphics[width=\columnwidth]{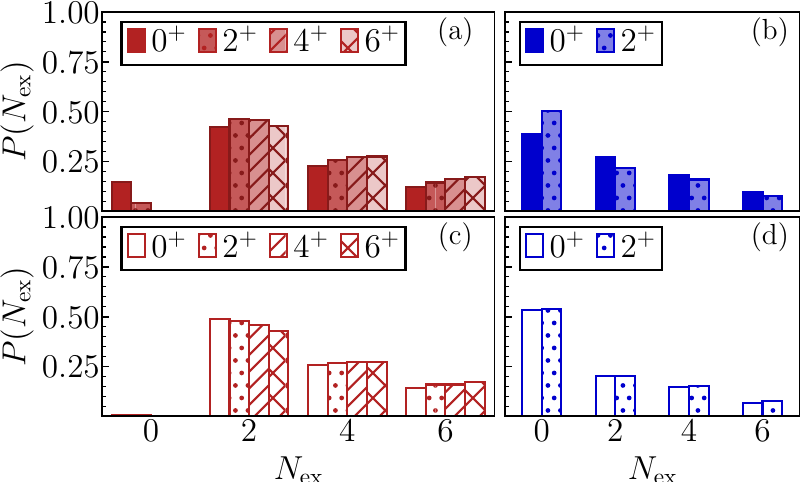}
  \end{center}
  \caption{\label{fig:Nex}Decomposition by $\Nex$ of wave functions of representative members of the (a) $K=0$ yrast band and (b) $K=0$ yrare band identified in Fig.~\ref{fig:network} (Sec.~\ref{sec:ncsm}). (c,d) Decomposition by $\Nex$ of pure wave functions (Sec.~\ref{sec:mixing}) of representative members of the same bands, respectively. Decomposition by $\Nex$ is trivially obtained by summing probability of harmonic oscillator configurations with given $\Nex$.
}
\end{figure}

Rotational states are characterized by a deformed intrinsic state rotating in the lab frame.
An intrinsic state which is rotationally symmetric about one of the principal axes is labeled by $K$, the projection of angular momentum $J$ onto the symmetry axis in the body-fixed frame. Rotational band members, \textit{i.e.}, states with the same intrinsic state but different angular momenta, are identifiable as states connected by $E2$ transitions enhanced relative to single particle estimates.  Band members have characteristic energies given by
\begin{equation}
  E(J)=E_0+\frac{\hbar^2}{2\mathcal{I}}J\,(J+1),
\label{eqn:rot-energy}
\end{equation}
where  $\mathcal{I}$ is the moment of inertia and $E_0$ is the energy intercept.

Rotational bands emerge in the low-lying spectrum of $\isotope[12]{Be}$, as shown in Fig.~\ref{fig:network}.  Excitation energies (symbols) are plotted versus $J$ on an axis that is scaled by $J(J+1)$ so that band members lie along a straight line.  Gray lines between states denote $E2$ transitions. Thickness and shading are proportional to the transition strength.  Three rotational bands are identified: two $K=0$ bands (red diamonds and blue circles) and a $K=2$ band (gold hexagons). Fits to the rotational energy formula are indicated by the dashed lines.   Excitation energies of the yrast $K=0$ band are in reasonable agreement with experimental values  (green lines)~\cite{npa-968-2017-71-Kelley} for the $0^+_1$, $2^+_1$ and (probable) $4^+_1$.\footnote{The state at 5.724~MeV has been classified as a $(4^+,2^+,3^-)$ state, but the $4^+$ assignment is preferred~\cite{npa-968-2017-71-Kelley}.}   The calculated bandhead of the yrare $K=0$ band lies just above the experimental $0_2^+$ state.  However, the calculated $2^+$ member of the yrare band over-predicts the possible $2^+$ state at 4.590(5) MeV by about 1~MeV.\footnote{The state at 4.580(5) MeV has been classified as a ($2^+,3^-$) state =~\cite{npa-968-2017-71-Kelley}.} The bandhead of the $K=2$ band is in good agreement with a probable $2^+$ state observed at 7.2(1)~MeV.

With access to the underlying calculated wave functions, we can  probe the structure of the  band members.  By decomposing the wave functions by number of excitation quanta $\Nex$, we demonstrate that all of the members of the yrast band (red diamonds) are intruder states.
 In the NCSM, wave functions are expanded in terms of configurations, \textit{i.e.}, distribution of particles over oscillator shells, with $\Nex$ up to some cutoff $\Nmax$.  We classify a state as ``normal'' if the largest single contribution to the wave function is $\Nex=0$ and as ``intruder'' otherwise.  To classify the bands in Fig.~\ref{fig:network}, wave functions of the band members are decomposed by $\Nex$ (Fig.~\ref{fig:Nex}).  Each of the states in the $K=0$ yrast band members have a largest contribution from $\Nex=2$ configurations [Fig.~\ref{fig:Nex}(a)].  Thus the states form an intruder band.   In contrast, the two states forming the yrare $K=0$ band  (blue circles) are normal states, \textit{i.e.}, with largest contribution from $\Nex=0$ configurations [Fig.~\ref{fig:Nex}(b)].   For the remainder of this paper, we will label the $K=0$ bands as intruder (\textit{i.e.}, $K=0_\intr^+$) and normal (\textit{i.e.}, $K=0_\nrm^+$).  The $K=2$ band (not shown in Fig.~\ref{fig:Nex}) is also an intruder band.

\begin{figure*}[tb]
  \begin{center}
    \includegraphics[width=.7\textwidth]{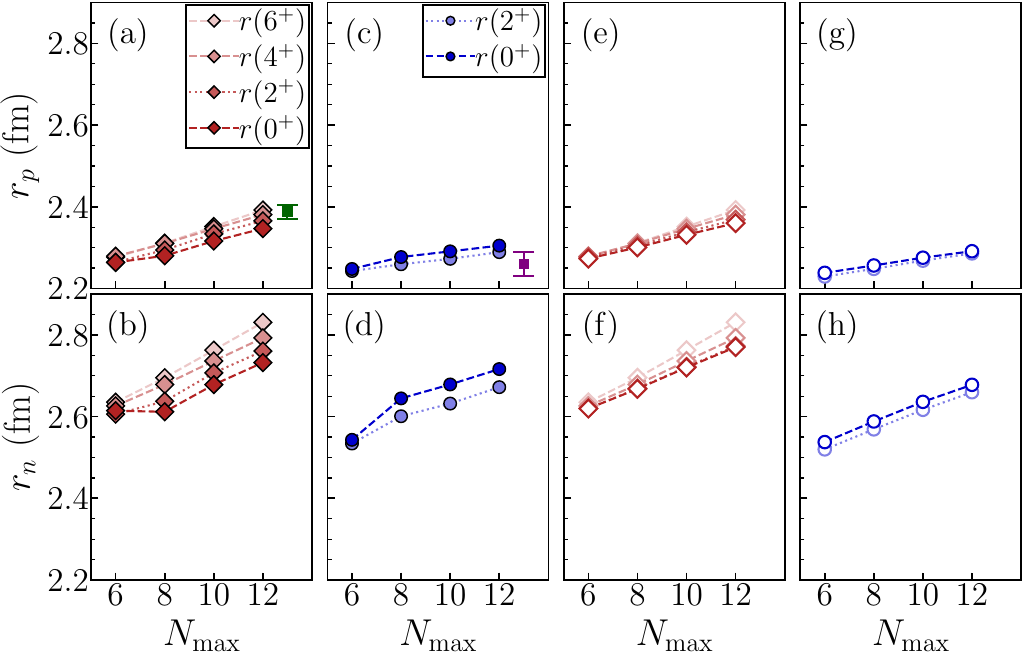}
  \end{center}

  \caption{\label{fig:radii} Proton and neutron radii with respect to $\Nmax$ of representative members of the (a,b) $K=0$  intruder, (c,d) normal, (e,f) pure intruder and  (g,h) pure normal bands.  All values shown for $\hbar\omega=15~\mathrm{MeV}.$  Experimental value for $r_p(0_1^+)$~\cite{npa-968-2017-71-Kelley} given by green square in panel (a).   Estimate for $r_p(0_2^+)$ inferred from known experimental quantities (Sec.~\ref{sec:mixing}) given by purple square in panel (c).  Note error bar on estimated $r_p(0_2^+)$ includes only experimental errors and does not account for error arising from assumptions made in estimation.}
\end{figure*}

Under the adiabatic assumption, the energy scale for rotational excitations is small compared to the energy scale for intrinsic excitations.  Thus the intrinsic structure is the same for all members of a band.\footnote{Here we assume that the members of the rotational bands can be factorized into an intrinsic wave function and a rotational wave function~\cite{Rowe2010}.   Expectation values of scalar operators which act as the identity on the rotational wave function can be identified as properties of the intrinsic state.  Such operators include $r^2$ as well as projection operators used in $\Nex$, $\grpsu{3}$ and spin decompositions presented in this paper. }
That the members of each band have similar $\Nex$ decompositions (Fig.~\ref{fig:Nex}) is approximately consistent with this assumption.

The intrinsic state is also assumed to have a well-defined quadrupole shape.  Here we consider the proton and neutron radii and intrinsic quadrupole moments $Q_0$, which characterize the quadrupole shape of the intrinsic state; these properties are related to the quadrupole deformation by $\beta\propto Q_0/\braket{r^2}$.   Since the $r^2$ operator is scalar, a radius calculated in the lab frame can immediately be identified with a radius in the body-fixed frame.  However, the intrinsic quadrupole moment can only be obtained indirectly from quadrupole moments and $B(E2)$ values.  For a symmetric rotor both these observables are proportional to the same intrinsic quadrupole moment, where the proportionality factors are given in terms of Clebsch-Gordan coefficients and angular momentum dimension factors~\cite{Rowe2010}:
\begin{equation}
\label{Q0-Q}
  Q(J)=\frac{3K^2-J(J+1)}{(J+1)(2J+3)}Q_0,
\end{equation}
and
\begin{equation}
\label{Q0-BE2}
B(E2;J_i\rightarrow J_f)=\frac{5}{16\pi}(J_iK20|J_fK)^2(eQ_0)^2,
\end{equation}
respectively.

For an ideal rotor, the radii of members of a rotational band are expected to be constant.  As shown in Fig.~\ref{fig:radii},  radii within each band are indeed similar in size, but values are not constant.  Although the values are not converged with respect to $\Nmax$,  proton and neutron radii in the intruder band members~[Fig.~\ref{fig:radii}(a,b)] have a clear dependence on angular momentum $J$. In the rotational framework,  the increase in radius with $J$ could be attributed to centrifugal stretching.  In the normal band~[Fig.~\ref{fig:radii}(c,d)], there is again an angular momentum dependence.  However, here, both the proton and neutron radii decrease with increasing $J$.

\begin{figure*}[tb]
  \begin{center}
    \includegraphics[width=.7\textwidth]{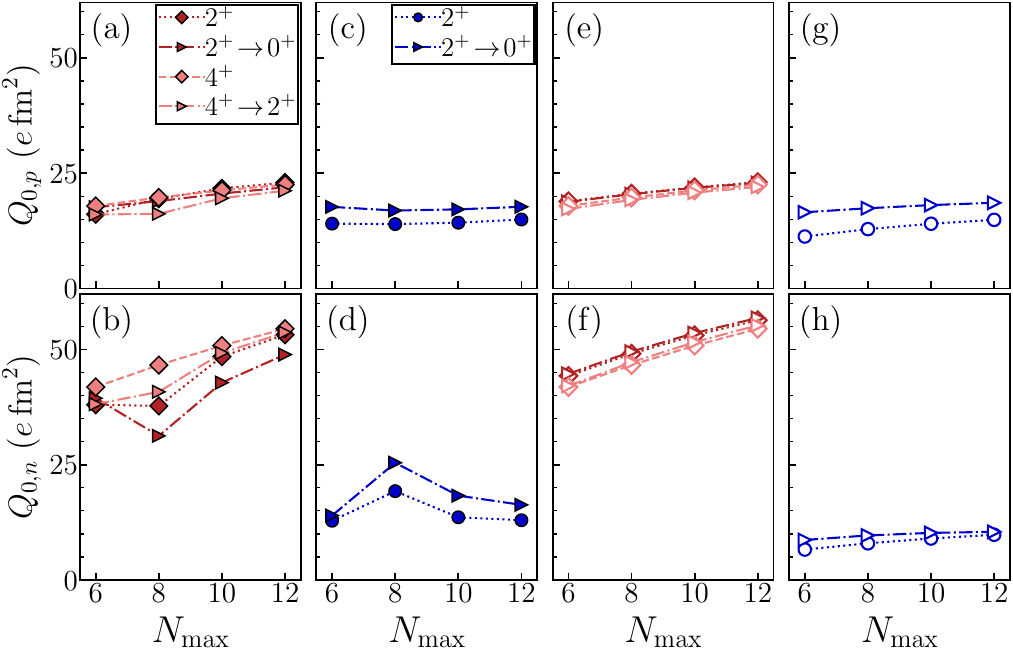}
  \end{center}
  \caption{\label{fig:Q0} Proton and neutron intrinsic quadrupole moments of the (a,b) $K=0$  intruder, (c,d) normal, (e,f) pure intruder, and  (g,h) pure normal bands with respect to $\Nmax$ as extracted from the calculated quadrupole moments and $B(E2)$ values.  All values shown for $\hbar\omega=15~\mathrm{MeV}.$}
\end{figure*}

Similarly, the  $Q_0$ extracted from the spectroscopic quadrupole moments and intraband $E2$ transitions should also be constant.  In Fig.~\ref{fig:Q0}, the (a) proton and (b) neutron intrinsic quadrupole moments of the intruder band are extracted from the quadrupole moments of the $2_\intr^+$ and $4^+_\intr$ states as well as the $2^+_\intr\rightarrow 0^+_\intr$ and $4^+_\intr\rightarrow2^+_\intr$ $B(E2)$  values.  For the normal band, (c) $Q_{0,p}$ and (d) $Q_{0,n}$ are obtained from the quadrupole moment of the $2^+_\nrm$  state and the $2^+_\nrm\rightarrow 0_\nrm^+$ $B(E2)$ value.   Though there are some small discrepancies in the values obtained for $Q_{0,p}$ in either band, the values overall appear to be comparable and thus consistent with rotor model expectations.
There is, however, a larger spread in the values extracted for $Q_{0,n}$ within each band, but the values do appear to be converging towards a more similar value by $\Nmax=12$.

While an approximate rotational picture emerges in the $\isotope[12]{Be}$ spectrum, the spread in radii and intrinsic quadrupole moments within each band suggests that the simple picture provided by the rotational model is incomplete.  An additional discrepancy is noted in the energies of the yrast band. As shown in Fig.~\ref{fig:network}, the  $0_\intr^+$ lies almost 1~MeV below the rotational energy (red dashed line) obtained by fitting \eqref{eqn:rot-energy} to the  $4^+$ and $6^+$ members of the intruder band.\footnote{Equivalently, if we were to fit the rotational energy formula to the $0^+$ and $2^+$ band members, the energies of the $4^+$ and $6^+$ band members would be well below the rotational energies.}
However, as we demonstrate in the following section, deviations from rotational expectations can largely be understood as a consequence of mixing~\cite{Casten2000} of same-$J$ members of the two rotational bands.

 \section{Band mixing}
\label{sec:mixing}
Deviations from rotational expectations can largely be understood as a consequence of mixing~\cite{Casten2000,prep-215-1992-101-Wood} between same-$J$ members of the two $K=0$ bands.   To gain insight into the mixing of these states, we apply a two-state mixing model in which we assume that the proton $E0$ transitions between pure intruder and normal states vanish.

Our assumption that the $E0$ transitions between the pure states vanish is motivated by the observation that the radii and quadrupole moments of the two bands differ.   As shown in Fig.~\ref{fig:radii}(a,c), the proton and neutron radii of  the $0^+$ and $2^+$ intruder band members are larger than those of the corresponding normal band members at each $\Nmax$.  Moreover, the $Q_{0,n}$ of the intruder band is significantly larger than that of the normal band, implying a significant difference in neutron quadrupole deformation and thus in the intrinsic shape.   In the limit where the intrinsic state is assumed to have definite shape (and is thus an eigenstate of $r^2$),  matrix elements of the $E0$ operator between bands with different intrinsic shapes must vanish.

Requiring the $E0$ transition between the pure states to vanish allows us to extract a mixing angle $\theta$, which can be used to extract values for energies, radii, and electromagnetic transitions and moments for the pure rotational bands.   The mixing angle between two states  $\ket{\psi_1}$ and $\ket{\psi_2}$ is then forced to be 
\begin{equation}
\tan2\theta=\frac{2\braket{\psi_1|\mathcal{M}({E}0)|\psi_2}}{\braket{\psi_2|\mathcal{M}({E}0)|\psi_2}-\braket{\psi_1|\mathcal{M}({E}0)|\psi_1}}.
\label{eqn:mixing}
\end{equation}
At $\Nmax=12$, the mixing angles between the two $0^+$ states and  two $2^+$ states are $\theta_{0^+}=26.3^\circ$ and $\theta_{2^+}=11.2^\circ$, respectively.

As shown in Fig.~\ref{fig:mixing_convergence}, the degree of mixing is highly $\Nmax$-dependent.  The fractions $P(\mathrm{pure})$ of the (left) $0_\intr^+$ and (right) $2^+_\intr$ state, coming from the pure intruder (red diamonds) and pure normal (blue circles) states, respectively, are shown in Fig.~\ref{fig:mixing_convergence}(a).   At $\Nmax=6$, the $0_\intr^+$ state is approximately 70\% pure intruder and 30\% pure normal.  Then at $\Nmax=8$, the state is nearly 50\% pure intruder and 50\% pure normal.  The fraction of the state which is pure normal then decreases with increasing $\Nmax$.  Note that, since the mixing of the states is symmetric, when the $0^+_\intr$ state is 70\% pure intruder and 30\% pure normal, the $0_\nrm^+$ is 70\% pure normal and 30\% pure intruder.  The $2^+$ states follow a similar evolution with $\Nmax$. By $\Nmax=12$, the $2^+$ states are almost entirely pure states.

Much of the $\Nmax$ dependence of the mixing is an  artifact of levels crossing as energies of the band members converge with $\Nmax$.  Though excitation energies within a band, and thus the moment of inertia, appear to be well-converged even at low $\Nmax$~\cite{epja-56-2020-120-Caprio}, different bands converge at different rates. The energies of the $0^+$ and $2^+$ members of the $K=0$ bands are shown in Fig.~\ref{fig:mixing_convergence}(b).  At $\Nmax=6$ the members of the normal band (blue circles) are lower in energy than those of the intruder band (red diamonds).  At $\Nmax=8$, where the states are maximally mixed, the pure states (open symbols) are nearly degenerate.  Within increasing $\Nmax$ the pure states move further apart in energy and the mixing correspondingly decreases.

Although both the $E0$ moment of the ground state (proportional to $[r_p(0_1^+)]^2$) and the $0_2^+\rightarrow 0_1^+$ $E0$ transition strength are measured~\cite{prl-108-2012-142501-Krieger,plb-654-2007-87-Shimoura}, an experimental value for the mixing angle $\thetaexp$ cannot be obtained as above, since the proton $E0$ moment of the $0_2^+$ state is not measured.  However, if we take the $2^+$ states to be essentially unmixed (which the NCSM calculation suggests is reasonable), we can deduce an approximate value for $\thetaexp$ from the known $E2$ transition strengths~\cite{npa-968-2017-71-Kelley}.  Specifically, $\tan\thetaexp \approx B(E2;2^+_1\rightarrow 0^+_2)/B(E2; 2^+_1\rightarrow0^+_1)$, and thus $\thetaexp\approx18^\circ\pm 2^\circ$.  The corresponding fraction of the physical states which come from the pure intruder state are shown in Fig.~\ref{fig:mixing_convergence}(a) (purple symbols labeled as Exp).  Although the calculated $\theta_{0^+}$ [and thus $P(\mathrm{pure})$ for the $0_\intr^+$ state] is not converged with respect to $\Nmax$, it appears to be converging towards a value that is reasonably consistent with $\thetaexp$.

Using the approximate experimental mixing angle, we can then extract an approximate value for $r_p(0_2^+)$ by inverting \eqref{eqn:mixing}.  Combining $\thetaexp$ with the measured ground state proton radius $r_p(0_1^+)$ and measured $E0$ transition between the $0^+$ states, we obtain the approximate value  $r_p(0_2^+)=2.26~\mathrm{fm}$, which is shown in Fig.~\ref{fig:radii}(c) (purple circle).   This approximate $r_p(0_2^+)$ is slightly smaller than the calculated $r_p(0_\nrm^+)$.

\begin{figure}[tb]
\begin{center}
\includegraphics[width=\columnwidth]{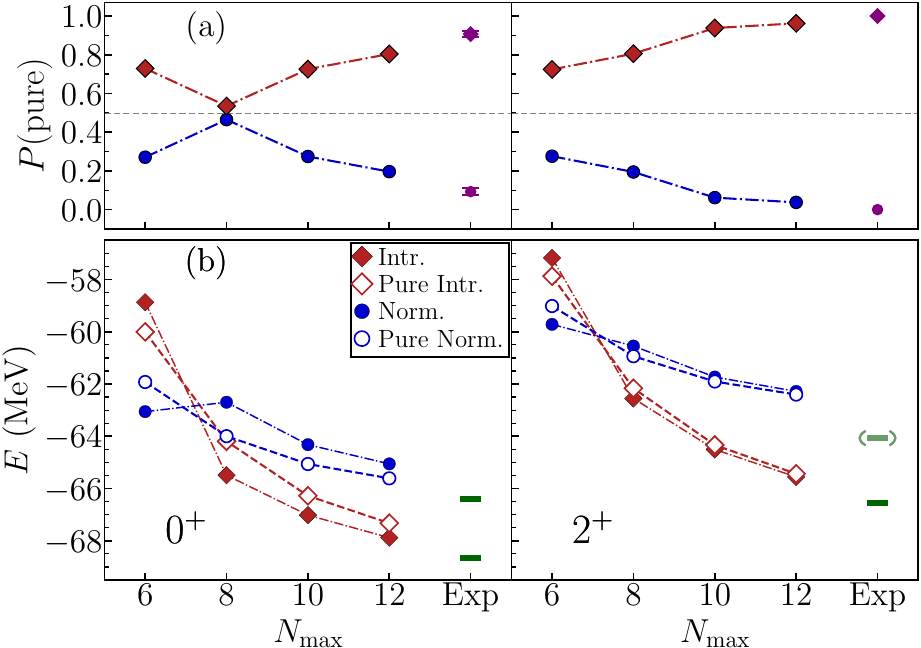}
\end{center}
\caption{\label{fig:mixing_convergence}(a) Fraction of the (left) $0^+_\intr$ and (right) $2^+_\intr$ wave function coming from the pure intruder state (red diamonds) and pure normal state (blue circles), given by $\cos^2\theta$ and $\sin^2\theta$, respectively.  Fractional contributions corresponding to an approximately deduced experimental mixing angle are indicated by purple symbols in the Exp. column. See Sec.~\ref{sec:mixing} for details. (b) Convergence of NCSM calculated energies of the  $0^+$ and $2^+$ band members (filled) and energies of the pure band members (open) with respect to $\Nmax$. Experimental values given by horizontal green line~\cite{cpc-45-2021-030003-Wang}. 
}
 \end{figure}

Like the $E0$ transition,  $E2$ transitions between pure band members with different intrinsic shapes are expected to vanish.  Calculated $E2$ transitions strengths are shown in Fig.~\ref{fig:transitions} between the (filled) mixed and (open) pure states.  Though not converged, the calculated $2^+_\intr\rightarrow 0^+_\intr$ and $2^+_\intr\rightarrow 0^+_\nrm$ transition strengths are reasonably consistent with experimental values (green squares).  The impact of the mixing on the interband $2^+\rightarrow 0^+$ transition for both the intruder and normal bands is small, in part because the matrix elements of $\mathcal{M}_p(E2)$ are similar in value at $\Nmax=6$ and 8 where the mixing is largest.  The effect on the interband transitions (purple x's) is more notable.  Transitions between the pure states with different intrinsic shape are expected to vanish.  As shown in Fig.~\ref{fig:transitions}, the transitions between the pure bands (open purple crosses)  are highly suppressed relative to the transitions between the mixed states (filled symbols).  Neutron interband transitions between pure states (not shown) are also highly suppressed. Note that the $2^+_2\rightarrow2^+_1$ transition strength is converging towards that between the pure states, providing further evidence that the $2^+$ states are nearly pure states by $\Nmax=12$.

\begin{figure*}[tb]
\begin{center}
\includegraphics[width=.6\textwidth]{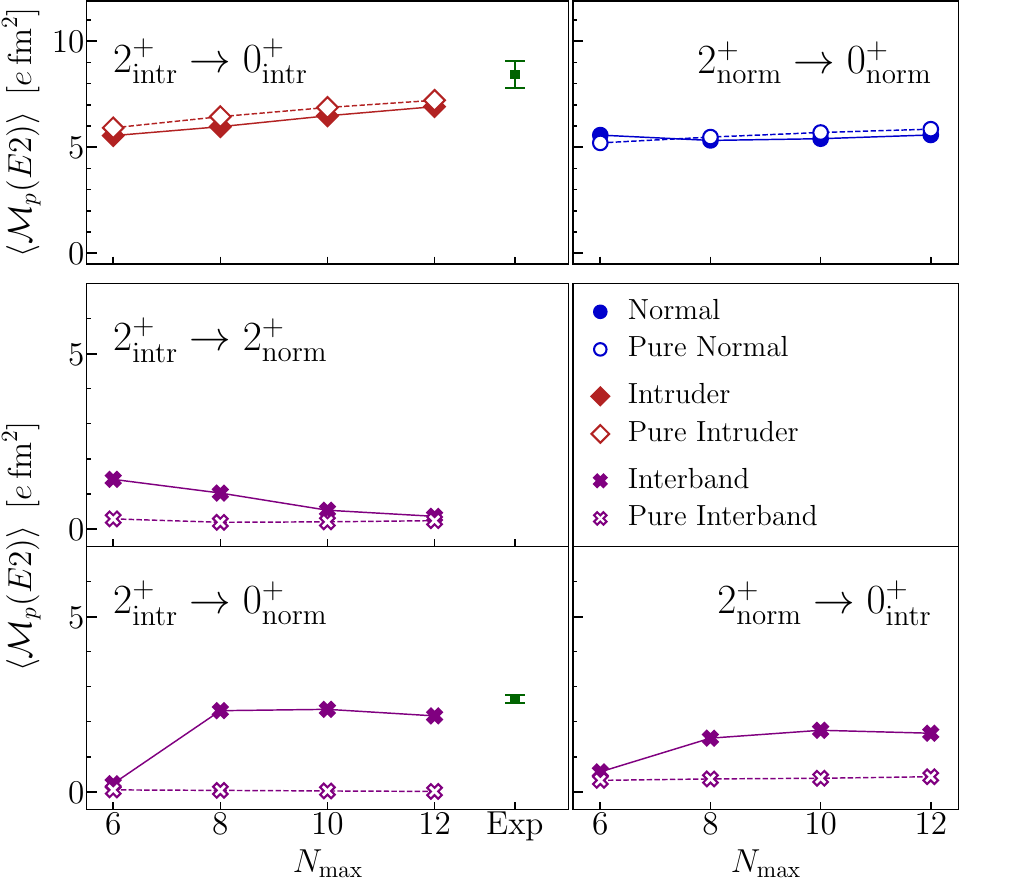}
\end{center}
\caption{\label{fig:transitions} Convergence with respect to $\Nmax$ of intraband (top) and interband (bottom) $E2$ transition strengths, shown as the magnitude of the reduced matrix element.  Experimental values (green squares) are given for $2_\intr^+\rightarrow 0_\intr^+$~\cite{plb-780-2018-227-Morse,plb-673-2009-179-Imai} and $2_\intr^+\rightarrow 0_\nrm^+$~\cite{plb-654-2007-87-Shimoura} transitions.}
\end{figure*}

With the two-state mixing picture established, we turn our attention to the impacts of mixing on the observables discussed above (Figs.~\ref{fig:network}-\ref{fig:mixing_convergence}).  Most of the discrepancies from rotational expectations noted in the previous section can be understood as resulting from mixing of the pure bands.

As noted in Sec.~\ref{sec:ncsm}, the energy of the $0_\intr^+$ state [Fig.~\ref{fig:network} (red filled symbols)] is pushed down relative to the rotational energies (red dashed line) obtained by fitting the rotational energy formula \eqref{eqn:rot-energy} to the $4^+$ and $6^+$ states.  This
difference in energy between the  $0^+_\intr$ state and the rotational prediction is largely a result of two-state mixing; level repulsion pushes the $0_\intr^+$  down in energy relative to the pure state, while pushing the $0^+_\nrm$ state up.  The energy of the pure $0^+_\intr$ state (red open symbol) is much closer to the rotational energy.

Notice that the level repulsion between the $0^+$ states creates an illusion regarding the moments of inertia of the bands.  If one naively interprets the energy difference between the  (mixed) $0^+$ and $2^+$ states as a measure of the rotational moment of inertia, then it appears as though the the bands have near identical moments of inertia. In contrast, the slope of the rotational energy fit to the pure normal states (blue dashed line) is more than 1.5 times larger than the slope of the rotational energy fit for the intruder band (red dashed line fitted to the $4^+$ and $6^+$ members), which translates into a moment of inertia which is 1.5 times smaller for the normal band than for the intruder band.

Both the radii and intrinsic quadrupole moments within each band are  more congruous, for the pure states, with the expectation that these properties be constant for all rotational band members.  As shown in Fig.~\ref{fig:radii}(e,g), the proton radii of both bands are nearly constant within a band.  Similarly, while the neutron radii of both bands [Fig.~\ref{fig:radii}(f,h)] do still have a $J$ dependence, the effect is much smaller.  The intrinsic quadrupole moments of the pure states [Fig.~\ref{fig:Q0}(e-h)] are also more consistent with the rotational expectations.  In particular, there is very little difference among the $Q_{0,n}$ values within each pure band. The exception to this pattern is the $Q_{0,p}$ of the pure normal band.  As shown in Fig.~\ref{fig:Q0}(g), the variation in $Q_{0,p}$ values extracted from the quadrupole moment $Q(2)$ and  $B(E2)$ of the pure states is larger.   The difference in extracted $Q_{0,p}$ may be an indication that the two pure normal states do not form a well-defined rotational band, even though the $E2$ transition between the states is enhanced.

The differences in radii between the pure intruder and pure normal states are also much more pronounced.  Most notable is the difference in neutron radii of the $0^+$ states:  whereas the neutron radii of the mixed $0^+$ states [Fig.~\ref{fig:radii}(b,d)] are very similar in value, the neutron radius of the pure $0^+_\intr$ state [Fig.~\ref{fig:radii}(f)] is significantly larger than that of the pure $0_\nrm^+$ state [Fig.~\ref{fig:radii}(h)]. By $\Nmax=12$ both the proton and neutron radii of the pure intruder band are more than $0.1~\mathrm{fm}$ larger than the corresponding radii of the pure normal band.  Note that, because the $E0$ transitions are assumed to vanish between pure states, the radii of the mixed states can be interpreted as weighted averages of the radii of the pure states, with weights determined by the mixing angle.

The large difference in $Q_{0,n}$ of the pure bands mirrors the difference in the moments of inertia of the two pure bands and thus suggests that the difference in moment of inertia reflects and underlying change in neutron intrinsic structure.  The $Q_{0n}$ of the pure normal band [Fig.~\ref{fig:Q0}(h)] is close to zero, as one would expect for a nucleus with a closed neutron shell.  This small $Q_{0,n}$ translates to a much smaller moment of inertia than that of the pure intruder band, as shown in Fig.~\ref{fig:network}.

The pure states are also more distinctly intruder and  normal in their $\Nex$ content.  Decompositions by $\Nex$ of the  pure states are obtained by first un-mixing the calculated wave functions of the $0^+$ and $2^+$ states.  The wave functions of the pure states can then be decomposed by $\Nex$ in the same manner as  the wave functions of the mixed states.   The $\Nex$ decompositions are shown in Fig.~\ref{fig:Nex} for the (c) pure intruder and (d) pure normal band members.   The pure $0_\intr^+$ and $2^+_\intr$ states look more ``intruder-like.'' The $\Nex=0$ contributions to the mixed intruder states [Fig.~\ref{fig:Nex}(a)] nearly vanish in the pure states.   In addition, the decompositions of the members within a band are all nearly identical.

Combining the rotational picture with two-state mixing provides a reasonable description of the low-lying positive parity spectrum of $\isotope[12]{Be}$.  In this section,  the assumption that the $E0$ transition vanishes between pure states is motivated by the assumption that the intrinsic state of the rotational bands is an approximate eigenstate of the $r^2$ operator.  In the following section, we demonstrate that vanishing transitions can also be motivated by the emergence of an approximate $\grpsu{3}$ symmetry, without assuming rotational structure or well defined intrinsic shape.  Though the $\grpsu{3}$ symmetry is only approximate, it does provide selection rules on the transition operators which support enforcing vanishing interband transitions.
 \section{Emergence of Elliott's $\grpsu{3}$ picture}
\label{sec:elliott-su3}
Elliott's $\grpsu{3}$ rotational framework provides a link between microscopic correlations and nuclear rotation and deformation.    In this framework, there is a rotational intrinsic state which has definite $\grpsu{3}$ symmetry with quantum numbers $(\lambda\mu)$~\cite{prsla-245-1958-128-Elliott,prsla-245-1958-562-Elliott}.  In the limit of large $\lambda$ and $\mu$, $(\lambda\mu)$ can be associated with the nuclear deformation parameters $\beta$ and $\gamma$, with larger values of $\lambda$ and $\mu$ corresponding to a more deformed intrinsic shape~\cite{ap-104-1977-134-Rosensteel,zpa-329-1988-33-Castanos}.  Microscopically,  each particle in a harmonic oscillator configuration has $\grpsu{3}$ quantum numbers $(\lambda\mu)=(N,0)$, where $N$ is the oscillator shell number for the given particle.  A many-body state with definite total $(\lambda\mu)$  can be obtained by coupling together $\grpsu{3}$ quantum numbers of  each particle according to $\grpsu{3}$ coupling rules.  This $\grpsu{3}$ state also has definite total spin $S$ obtained by coupling the spins of each particle together according to angular momentum coupling rules.

Elliott's framework captures the competition between shell structure and correlation energy.
In this framework, the model Hamiltonian is typically given by $\hat{H}=\hat{H}_0-\chi \mathcal{\hat{Q}}\cdot\mathcal{\hat{Q}}$, where $H_0$ is the harmonic oscillator Hamiltonian, $\chi$ is a strength parameter, and $\mathcal{\hat{Q}}$ is an $\grpsu{3}$ generator which is closely related to the physical (mass) quadrupole operator but cannot move a nucleon between oscillator shells~\cite{prsla-272-1963-557-Elliott,anp-1-67-1968-harvey,ppnp-120-2021-103866-Nowacki}. The first term, $H_0$, gives rise to shell structure, while the (negative) quadrupole-quadrupole term gives preference to more deformed nuclear states.

Traditionally, Elliott's model was restricted to the shell model valence space, which maps implicitly onto the $\Nex=0$ subspace.  However, such a picture cannot describe intruder states.  Thus, to describe $\isotope[12]{Be}$, we extend the framework to include $\grpsu{3}$ states in either the $\Nex=0$ or the $\Nex=2$ subspaces.  Within each of these subspaces, the largest deformations correspond to $\grpsu{3}$ many-body states with quantum numbers $\Nex(\lambda\mu)S=0(2,0)0$ and $2(6,2)0$, respectively.   Qualitatively, the very large deformation associated with a $2(6,2)0$ many-body state, as compared with that of a $0(2,0)0$ state, overcomes the (positive) harmonic oscillator energy required to excite particles out of the valence space, bringing the $J=0,2$ intruder band members below the normal states.

The Elliott model Hamiltonian gives rise to rotational bands.  The $\mathcal{\hat{Q}}\cdot\mathcal{\hat{Q}}$ term in the Hamiltonian can be re-expressed in terms of the $\grpsu{3}$ Casimir operator $\hat{C}_{\mathrm{\grpsu{3}}}$ and the orbital angular momentum operator $\hat{L}^2$.  The Hamiltonian then becomes $\hat{H}=[\hat{H}_0-\chi \hat{C}_{\mathrm{\grpsu{3}}}]+3\chi \hat{L}^2$, where the eigenvalue of $H_0-\chi\hat{C}_{\mathrm{\grpsu{3}}}$ corresponds to $E_0$ in \eqref{eqn:rot-energy}\footnote{The eigenvalue of the $\grpsu{3}$ Casimir operator for states with definite $(\lambda\mu)$ is given by $\braket{\hat{C}_{\mathrm{\grpsu{3}}}}=4(\lambda^2+\lambda\mu+\mu^2+3(\lambda+\mu))$.} and the eigenvalue of $3\chi L^2$ corresponds (for $S=0$) to the $J(J+1)$ term in~\eqref{eqn:rot-energy}.  Unlike the simplified rotational picture presented in Sec.~\ref{sec:ncsm}, the $\grpsu{3}$ intrinsic states are not presumed to be symmetric about any of the principal axes.  Thus more than one band, with different $K$ quantum numbers, can be projected out from the same $\grpsu{3}$ intrinsic state.

For $\isotope[12]{Be}$, the rotational bands identified in Fig.~\ref{fig:network} are qualitatively consistent with the rotational bands expected in Elliott's framework.  The band projected out from the most deformed $\grpsu{3}$ state in the $\Nex=0$ space [$\Nex(\lambda\mu)S=0(2,0)0$] has members with $J=0$ and $2$, which qualitatively matches the states appearing in the $K=0$ normal band in Fig.~\ref{fig:network}.   An intrinsic state with quantum numbers $\Nex(\lambda\mu)S=2(6,2)0$ projects out onto two bands, a $K=0$ band with $J=0,2,...,8$ members and a $K=2$ band with $J=2,3,...,7$ members.  The angular momentum of the states appearing in these bands are consistent with the $K=0$ and $K=2$ intruder bands identified in Fig.~\ref{fig:network}.  Because an $\grpsu{3}$ intrinsic state has definite $\Nex$, the resulting bands necessarily terminate at or below the maximum $J$ allowed within the $\Nex$ subspace.  For example, the maximum $J$ allowed in the $\Nex=0$ space is $J=2$, and thus the normal band cannot extend past $J=2$.

\begin{figure}[tb]
\begin{center}
\includegraphics[width=\columnwidth]{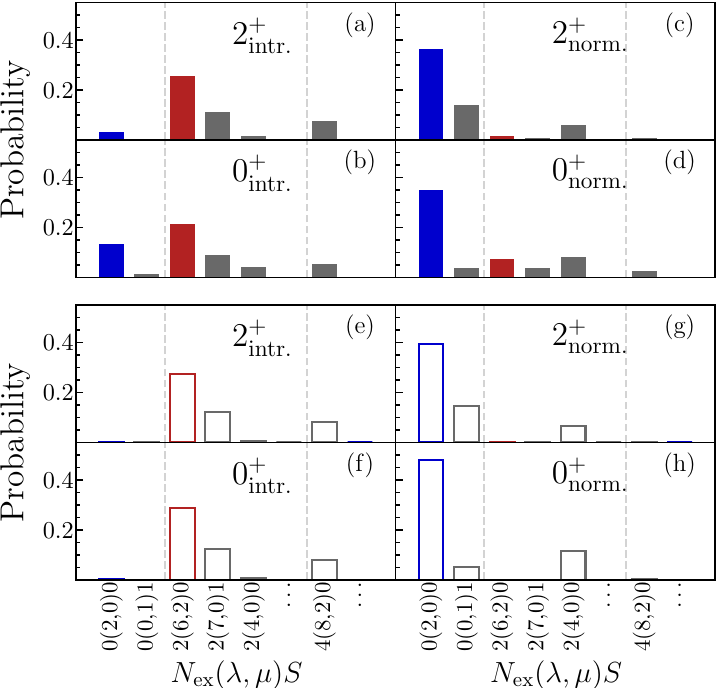}
\end{center}
\caption{
\label{fig:su3-Nmax12}
Decompositions of wave functions into contributions labeled by $\Nex(\lambda\mu)S$ for the (a,b) intruder,  (c,d) normal,  (e,f) pure intruder, and (g,h) pure normal $0^+$ and $2^+$ states. Only those subspaces contributing $\ge 5\%$ to at least one of the states are shown.
}
\end{figure}

To identify the $\grpsu{3}$ content of the intrinsic states of the bands, we decompose the wave functions of the band members into contributions from subspaces with definite $\Nex(\lambda\mu)S$.~\footnote{Here we use the ``Lanczos trick'' to decompose the wave function.  See, \textit{e.g.}, Refs.~\cite{Whitehead1980,gueorguiev2000:fp-su3-breaking,johnson2015:spin-orbit,herrera2017:cr-backbending-qds-su3-su4,jpg-48-2021-075102-Zbikowski,bjp-49-2022-057066-Caprio}.}    Decompositions of the $0^+$ and $2^+$ band members into different $\Nex(\lambda\mu)S$ contributions are shown in Fig.~\ref{fig:su3-Nmax12}.  Only subspaces contributing $\ge 5\%$ to at least one state are shown.  As expected, for both of the intruder states [Fig.~\ref{fig:su3-Nmax12}(a,b)], the largest contribution is from the subspace labeled by $\Nex(\lambda,\mu)S=2(6,2)0$.  (This is also true for the $K=2$ band, not shown.)  Also as expected, the largest $\Nex(\lambda,\mu)S$ contribution to the normal $K=0$ band members [Fig.~\ref{fig:su3-Nmax12}(c,d)] comes from the $0(2,0)0$ subspace.  These decompositions confirm that the bands shown in Fig.~\ref{fig:network} are consistent with those expected in the Elliott rotational framework.

Of course, Elliott's $\grpsu{3}$ symmetry is only approximate~\cite{prl-111-2013-252501-Dytrych,cpc-207-2016-202-Dytrych,McCoy2018,prl-125-2020-102505-McCoy,epja-56-2020-120-Caprio,jpg-48-2021-075102-Zbikowski,bjp-49-2022-057066-Caprio}.  The largest single $\Nex(\lambda,\mu)S$ contribution  in both bands is less than $50\%$, with the remaining probability fragmented over many other subspaces with different number of $\Nex$.   However, a substantial part of the fragmentation is due to the two-state mixing; Fig.~\ref{fig:su3-Nmax12}(e-h) shows the decompositions for the pure states. In the pure intruder band [Fig.~\ref{fig:su3-Nmax12}(e,f)], the decompositions of the $0_\intr^+$ and $2_\intr^+$ states are nearly identical.  For the normal band [Fig.~\ref{fig:su3-Nmax12}(g,h)], contributions arise with the same quantum numbers, but their relative magnitudes vary.  Notably, the $2^+_\nrm$ state [Fig.~\ref{fig:su3-Nmax12}(g)] has a more significant $S=1$ contribution from states labeled by $0(0,1)1$, which may indicate a weakening of an underlying $2\alpha$ cluster structure. This difference in decompositions also provides context for the differences in the $Q_{0,p}$ values extracted for the pure normal band in Fig.~\ref{fig:Q0}(g). We note that, in both the pure normal and pure intruder states, much of the remaining fragmentation over $\Nex$ comes from subspaces labeled by quantum numbers which would be consistent with symplectic excitations~\cite{jpg-35-2008-095101-Dytrych,prl-98-2007-162503-Dytrych,prc-76-2007-014315-Dytrych,prl-125-2020-102505-McCoy,jpg-48-2021-075102-Zbikowski,prl-124-2020-042501-Dytrych,epj-229-2429-2020-Launey}.

Despite  $\grpsu{3}$ symmetry breaking, there is no significant overlap in the distributions for the pure intruder band and for the pure normal band.
The stark difference in $\grpsu{3}$ content of the two bands provides a more microscopically based interpretation of why interband transitions between the pure bands vanish.  Both the $E0$ and $E2$ transition operators can be expressed as a linear combination of $\grpsu{3}$ tensors, namely $[b^\dagger\times b]^{0(1,1)0}$, $[b^\dagger\times b^\dagger]^{2(2,0)0}$, $[b\times b]^{-2(0,2)0}$, and $[b^\dagger \times b]^{0(0,0)}$~\cite{jpg-47-2020-122001-Caprio,McCoy2018}, where $b^\dagger$ and $b$ are the boson creation and annihilation operators.  From $\grpsu{3}$ and spin selection rules on each tensor term, it can be shown that the matrix elements of either operator vanish between any subspace that contributes $\ge 5\%$ to the pure intruder band and any that contributes $\ge 5\%$ to the pure normal states.

\section{Conclusion}
\label{sec:conclusion}

In this work we have investigated the underlying structure of the low-lying positive parity states of $\isotope[12]{Be}$.  As we demonstrate, the intruder nature of the lowest lying state emerges \emph{ab initio} in the NCSM framework, without any assumptions of, \textit{e.g.}, shell structure, clustering, or symmetry.  With the calculated energies, radii and electromagnetic transitions in reasonable agreement with experiment, the calculated wave functions can now be used to probe the underlying structure of the low lying spectrum of $\isotope[12]{Be}$.

Within the \emph{ab initio} calculated spectrum for $\isotope[12]{Be}$, signatures of nuclear rotations emerge.  Rotational bands are identified as states connected by enhanced $E2$ transitions with energies approximately consistent with characteristic rotational energies.  In particular, a $K=0$ intruder band built on the ground state and a normal $K=0$ band built on the first excited $0^+$ appear.  However, a simple symmetric rotor model is insufficient to describe the $\isotope[12]{Be}$ spectrum.  The $0_\intr^+$ energy deviates from the rotational $J(J+1)$ energy relation, while radii and intrinsic quadrupole moments are inconsistent within each band. Decompositions of band members by $\Nex$ and $\grpsu{3}$ symmetry also highlight inconsistencies in the intrinsic structure within each band.

Band mixing can explain the discrepancies between the NCSM-calculated observables and the rotational picture.  The low-lying spectrum can thus be described in terms of mixing between pure bands with very different intrinsic structure, namely an intruder band and a normal band. By assuming that the proton $E0$ transitions between the pure bands vanish, we deduce a mixing angle and use it to extract properties of the pure bands from the NCSM-calculated observables.  The (extracted) observables, e.g., radii and intrinsic quadrupole moments, associated with the pure bands, as well as the energy of the pure $0^+_\intr$ state,  are significantly more consistent with rotational model expectations.  Moreover, $\Nex$ and $\grpsu{3}$ symmetry decompositions are more constant within each band.

Both  of the $K=0$ bands as well as the $K=2$ intruder band exhibit an approximate $\grpsu{3}$ symmetry. Within each band, the largest $\grpsu{3}$ contribution comes from the same $\grpsu{3}$ subspace, notably the $\grpsu{3}$ subspace associated with the largest deformation in the corresponding $\Nex$ subspace.  Moreover, the angular momenta of the band members are exactly those expected in Elliott's rotational model for an intrinsic state with quantum numbers corresponding to that largest contributing symmetry subspace.   Although the $\grpsu{3}$ symmetry is only approximate, the pure states have notable contributions from only a few $\grpsu{3}$ subspaces.  Much of the apparent fragmentation of $\grpsu{3}$ is instead a result of the two-state mixing.

The mixing framework applied in this work assumes $E0$ transitions between the pure bands vanish.  This assumption is typically motivated by the argument that the $E0$ operator cannot connect states with very different intrinsic shape. In light nuclei intrinsic shape is often not well defined. However, the stark difference in the $\grpsu{3}$ content of the pure bands provides a microscopic explanation for vanishing $E0$ interband transitions: selection rules forbid $E0$ transitions between any of the $\grpsu{3}$ subspaces  contributing significantly to the intruder band members and any of the $\grpsu{3}$ subspaces contributing significantly to the normal band members.

Thus, a remarkably simple picture emerges from the \emph{ab initio} calculated spectrum,  for which the only input was the inter-nucleon interaction.  The low lying spectrum of $\isotope[12]{Be}$ can be described as mixing of a $K=0$ intruder band and a $K=0$ normal band with very different intrinsic structure.

 \appendix

\section*{Acknowledgements}
We thank B.~P.~Kay,  R.~B.~Wiringa, M.~B.~Colianni, and T.~E.~Corpuz for useful discussions and feedback on this manuscript.
This material is based upon work supported by the U.S.~Department of Energy,
Office of Science, under Award Numbers
DE-FG02-00ER41132, DE-SC0021027, DE-SC0013617 (FRIB Theory Alliance), DE-AC02-06CH11357, DE-FG02-95ER40934, and DE-SC0023495 (SciDAC5/NUCLEI). An award of computer time was provided by the Innovative and Novel Computational
Impact on Theory and Experiment (INCITE) program.  This research used resources
of the National Energy Research Scientific Computing Center (NERSC), a DOE
Office of Science User Facility supported by the Office of Science of the
U.S.~Department of Energy under Contract No.~DE-AC02-05CH11231, using NERSC
award NP-ERCAP0023497, and of the Argonne Leadership Computing Facility (ALCF),
a DOE Office of Science User Facility supported by the Office of Science of the
U.S.~Department of Energy under Contract DE-AC02-06CH11357.

\bibliographystyle{elsarticle-num}
\biboptions{sort&compress}

\end{document}